\documentclass[12pt]{article}

\usepackage{graphicx}
\usepackage{dsfont}
\usepackage{epstopdf}
\begin{document}
\thispagestyle{empty}

\title{Velocity measurements in General Relativity revisited}

\author{{\bf Miguel A. Oliveira} $^{ \footnote{Email address:masm.oliveira@gmail.com}}$}

\date{\today}
\maketitle

\baselineskip 24pt

\newpage
\thispagestyle{empty}
\begin{abstract}

\baselineskip 24pt

In this work we generalize an earlier treatment of the measurements of velocities at the event horizon of a black hole. This is intended as a pedagogical exercise as well as one more contribution to the resolution of some unphysical interpretations related to velocity measurements by generalized observers. We now use a more general metric and, non-geodesic observer sets to show that the velocity of a test particle at the event horizon is less than the speed of light.

\end{abstract}

\vspace{1.2cm}

\newpage

\section{Introduction}

\baselineskip 20pt

Black holes have gone in recent years, from hypothetical possibilities to actual astrophysical objects, detected in a number of circumstances, and by many observational methods. Their importance has grown with the discovery of the existence of Supermassive Black Holes (SMBH) at the centres of Active Galactic Nuclei (AGN).  However, some unphysical results have found there way into the literature, causing widespread confusion an some misconceptions. One of these, is the notion that \emph{for all observers}, a test particle will reach the event horizon at the speed of light. This statement, has appeared in some very well known books like~\cite{Landau,Zeldovich,Frolov,Shapiro}. This problem in particular was analyzed first by Tereno~\cite{Tereno1}, and later by Crawford and Tereno~\cite{Crawford1}. In the later reference, these authors discussed the frequently overlooked, difference between `coordinate charts' and `frames of reference' and they gave a prescription for the velocity of a test particle at a spcetime point, relative to some generalized observer at that point. Let $t^a$ be the 4-velocity of the test particle, there is a natural way for an observer with 4-velocity $u^a$ to define the velocity $v$ at which that test particle passes him (in a spacetime point $p$)\cite{Crawford1}:
\begin{equation}\label{vel-generic}
    v^2=\frac{\left(g_{ab}+u_au_b\right)t^at^b}{(u_at^a)^2}=1-\frac{1}{(u_at^a)^2}\,.
\end{equation}
Where in the last equality the fact that $u_au^a=-1$, was used.

Note that $t^a$ can always be cast in the form:
\begin{equation}
    t^a=t_{\|}+t_{\bot}=\lambda u^a + \ell^a\quad\mbox{where,}\quad \ell^au_a=0,.
\end{equation}
If $t^a$ is timelike $t^at_a=-\lambda^2+|\ell^2|<0$, and since we have that $t^a$ and $u^a$ are future pointing, $\lambda=-u_at^a>0$ and, $|\ell|<\lambda=(|\ell|^2+1)^{1/2}$. Therefore, using this last equation with the last equality in eq.~(\ref{vel-generic}) we find that $v^2<1$ under these conditions. And this concludes the coordinate free proof in \cite{Crawford1} that the measured velocities of material particles must be less the the speed of light.

Following this proof, they applied the above formula to static and radial observers in the case of Schwarzschild spacetime.

In this work, in section 2, we use a slightly more general metric, and apply the same prescription to static, radial observers, to circular accelerated observers and finally to spiral infalling observes. In section 3, we particularize the the case of the Schwarzschild solution. Finally, in section 4 we use another particular solution of the Einstein equations that has become popular in the literature, the de Sitter Schwarzschild solution.

The results here are in agreement, and extend those in \cite{Crawford1}. We find that static and circular observers cannot exit at the event horizon. Therefore, the fact that the limit of some expressions for the velocity is one at the event horizon is according to these results unphysical.

\section{Velocity measurements spherisymmetric geometry }
General static spherisymmetric geometry:
\begin{equation}\label{metric}
    ds^2=-V^2(r)dt^2+\eta^2(r)dr^2+r^2d\Omega^2
\end{equation}
Where, we use the shorthand $d\Omega^2=d\vartheta^{\,2}+\sin^2(\vartheta)d\varphi{\,^2}$, for the unit 2-sphere line element. We have a horizon at $r_{horz}$ if $V(r_{horz})^2=0$.
The Lagrangian is:
\begin{equation}\label{lagrangian}
    2\mathcal{L}=-V^2\dot{t}^2+r^2\omega^2+\eta^2\dot{r}^2\,.
\end{equation}
Where, $\dot{\ }\equiv\frac{d\   }{d\tau} $ and, $\omega=\dot{\varphi}$.
Since, $\mathcal{L}$ is not a function of $t$ and $\varphi$, these are cyclical variables and therefore:
\begin{eqnarray}
  \label{energy}-\frac{\partial\mathcal{L}}{\partial\dot{t}} &=& V^2\dot{t}\equiv E \\
  \label{angular}\frac{\partial\mathcal{L}}{\partial\dot{\varphi}} &=& r^2 \omega\equiv L
\end{eqnarray}
And these quantities are constants of the motion.
Using $E$ and $L$ we can write:
\begin{equation}\label{lag-EL}
    2\mathcal{L}=\frac{E^2}{V^2}-\frac{L^2}{r^2}-\eta^2\dot{r}^2\,,
\end{equation}
and therefore deduce the equation of motion:
\begin{equation}\label{eq-motion}
    \frac{1}{2}(\dot{r})^2=\frac{1}{2}\frac{E^2}{V^2\eta^2}-\frac{1}{2}\frac{L^2}{r^2\eta^2}-\frac{\mathcal{L}}{\eta^2}=\frac{1}{2}\frac{E^2}{V^2\eta^2}-\mathcal{V}(r)\,,
\end{equation}
where $\mathcal{V}(r)$, is analogous to the potential energy function in Newtonian one dimensional mechanics.

For radial geodesics we have:
\begin{equation}\label{tau}
    -d\tau^2=-V^2+´\eta^2dr^2\,,
\end{equation}
therefore,
\begin{equation}\label{v}
    d\tau^2=V^2dt^2\left(1-v^2\right)\,,
\end{equation}
where:
\begin{equation}\label{v2}
    v^2=\left(\frac{\eta}{V}\right)^2\left(\frac{dr}{dt}\right)^2\,.
\end{equation}
For a temporal geodesic we  have $\mathcal{L}=-\frac{1}{2}$, therefore from~(\ref{lagrangian}),
\begin{equation}\label{lag-r}
    1=V^2\dot{t}^2-\eta^2\dot{r}=\frac{E^2}{V^2}-\eta^2\left(\frac{dr}{d\tau}\right)^2\,,
\end{equation}
whence,
\begin{equation}\label{drdtau}
    \left(\frac{dr}{d\tau}\right)^2=\frac{E^2}{V^2\eta^2}-\frac{1}{\eta^2}\,.
\end{equation}
It can easily be verified that, $\frac{E^2v^2}{V^2\eta^2}=\left(\frac{dr}{d\tau}\right)^2$, therefore form eq.~(\ref{drdtau}) we have:
\begin{equation}\label{initial}
    E^2=\frac{V^2}{1-v^2}=\frac{V(R)^2}{1-v_0^2}\,,
\end{equation}
were $R$ and $v_0$ are the initial conditions and, in the last equality we used the fact that $E$ is a constant of the motion.

We have from eqs.(\ref{energy}) and (\ref{drdtau})  the following 4-velocity for a test particle in free fall:
\begin{equation}\label{4vfree}
    t^a=\left(
            \frac{dt}{d\tau}\,,  \frac{dr}{d\tau}\,,  0\,,  0
        \right)=\left(
                    \frac{E}{V^2}\,,  -\sqrt{\frac{E^2}{\eta^2V^2}-\frac{1}{\eta^2}}\,,  0\,,  0
                \right)\,.
\end{equation}
\subsection{Static observers}
For a static, (and therefore non geodesic) observer, we have the 4-velocity:
\begin{equation}\label{4vstatic}
    u^a=\sqrt{g_{00}}\delta_{a0}\,,
\end{equation}
because $u^au_a=-1$.

Using the general formula found by \cite{Crawford1}:
\begin{eqnarray}\label{vstat}
  \nonumber v^2 &=& 1-\frac{1}{(u^at_a)^2} \\
   &=& 1-\frac{V^2}{E^2}\,.
\end{eqnarray}
We have, that $v\rightarrow1$ as $V\rightarrow0$. We must keep in mind however, that such an observer is accelerated and therefore can not be at the event horizon.
\subsection{Radial observers}
For a radially infalling observer we have:
 \begin{equation}\label{radial-vs}
    t^a=\left(
                    \frac{E_1}{V^2}\,,  -\sqrt{\frac{E^2_1}{\eta^2V^2}-\frac{1}{\eta^2}}\,,  0\,,  0
                \right)\quad
                u^a=\left(
                    \frac{E_2}{V^2}\,,  -\sqrt{\frac{E^2_2}{\eta^2V^2}-\frac{1}{\eta^2}}\,,  0\,,  0
                \right)\,.
 \end{equation}
 And for the velocity measured by this (now geodesic) observer we have:
 \begin{equation}\label{vradial}
    v^2=1-\frac{V^4}{E^2_1E^2_2\left[1-\sqrt{1-\frac{V^2}{E^2_1}}\sqrt{1-\frac{V^2}{E^2_2}}\right]^2}\,.
 \end{equation}
 Now, if $V\rightarrow0$ we have an indeterminacy, since $v^2\rightarrow1-\frac{0}{0}$.
\subsection{Circular observers}
Circular observers can be geodesic or accelerated depending on the value of the coordinate $r$, see for example \cite{Israel}.

The velocity of a unit test mass in a circular orbit is:
\begin{equation}
    \frac{v^2}{r}=\frac{V'(r)}{V(r)}\,.
\end{equation}
If we take $r_{\gamma}$ to be the r-coordinate of a photon circulating in a geodesic orbit, we have:
\begin{equation}\label{rph}
    \frac{1}{r_{\gamma}}=\frac{V'(r_{\gamma})}{V(r_{\gamma})}\,.
\end{equation}
For $r<r_{\gamma}$, orbits are accelerated. The magnitude of the force needed to hold a unit test mass in a circular orbit is:
\begin{equation}\label{force}
    F=\frac{V'}{\eta V}\left\{1-\omega^2r\left(\frac{V}{V'}-r\right) \right\}=F_{stat}\left\{1-\omega^2r\left(\frac{V}{V'}-r\right) \right\}\,.
\end{equation}
Here, $F_{stat}=F|_{\omega=0}$, showing that a force needs to be exerted to hold a unit test mass stationary.

For these observers the 4-velocity becomes:
\begin{equation}\label{circular-4v}
    u^a=\left(\frac{dt}{d\tau}\,,0\,,0\,,\frac{d\varphi}{d\tau}\right)=\left(\frac{(1+\omega^2r^2)^{1/2}}{V}\,,0\,,0\,,\omega\right)\,.
\end{equation}
Where in the last equality we used $u_au^a=-1$. Using~(\ref{4vfree}), for the test particle and the formula in \cite{Crawford1} we have:
\begin{equation}\label{vcirc-test}
    v^2=1-\frac{V^2}{E^2(1+´\omega^2r^2)}\,.
\end{equation}
 Equations~(\ref{circular-4v})~and~(\ref{vcirc-test}) reduce to eqs.~(\ref{4vstatic})~and~(\ref{vstat}) respectively in the case $\omega=0$. For these accelerated observers the limit $V\rightarrow0$ is also $v\rightarrow1$, exactly because they are not geodesic and so no such observer will exist at the event horizon.
\subsection{Spiral infalling observers}
As we previously mentioned, no circular geodesic orbits exist for $r<r_{\gamma}$. However, spiral infalling geodesic observers exist and they can cross the path of the test particle and thus perform the velocity measurement.

For these observers, eqs.~(\ref{energy})~and~(\ref{angular}) are still valid, and eq.~(\ref{drdtau}) becomes:
\begin{equation}\label{drdtauomega}
    \left(\frac{dr}{d\tau}\right)^2=\frac{E^2}{V^2\eta^2}-\frac{L^2}{\eta^2r^2}-\frac{1}{\eta^2}\,.
\end{equation}
So from these last equations, we have for the 4-velocities in this case
\begin{equation}\label{4v-spiral}
    t^a=\left(
                    \frac{E_1}{V^2}\,,  -\sqrt{\frac{E^2_1}{\eta^2V^2}-\frac{1}{\eta^2}}\,,  0\,,  0
                \right)\quad
                u^a=\left(
                    \frac{E_2}{V^2}\,,  -\sqrt{\frac{E^2_2}{\eta^2V^2}-\frac{L^2}{\eta^2r^2}-\frac{1}{\eta^2}}\,,  0\,,  0
                \right)\,.
\end{equation}
And, the velocity measurement yields:
\begin{equation}\label{v-spiral}
    v^2=1-\frac{V^4}{E^2_1E^2_2\left[1-\sqrt{1-\frac{V^2}{E^2_1}}\sqrt{1-\left(\frac{L^2}{r^2}+1\right)\frac{V^2}{E^2_2}}\right]^2}\,.
\end{equation}
Which reduces to~(\ref{vradial}) in the limit $L\rightarrow0$. Also, in this case the limit $V\rightarrow0$ is indeterminate since, $v^2\rightarrow1-\frac{0}{0}$.
\section{The Schwarzschild case}
In the case of the Schwarzschild geometry we have:
\begin{equation}\label{Schw}
    V^2(r)=\eta(r)^{-2}=\alpha(r)=\left(1-\frac{2m}{r}\right)\,.
\end{equation}
The photon radius is $r_{\gamma}=3m$ and, the event horizon is at $r_{s}=2m$. Therefore, no circular geodesic orbits exist for $2m<r<3m$.
The force needed to hold a stationary test particle reduces to
$$F_{stat}=V'(r)=\frac{\frac{m}{r^2}}{\sqrt{1-\frac{2m}{r}}}=\frac{m}{r^2\alpha^{1/2}}\,,$$
and eq.~(\ref{force}) becomes:
\begin{equation}\label{}
    F=F_{stat}\left\{1-\omega^2r\left(\frac{r^2\alpha}{m}-r\right) \right\}\,.
\end{equation}
\begin{figure}
\centering
\includegraphics[width=8cm]{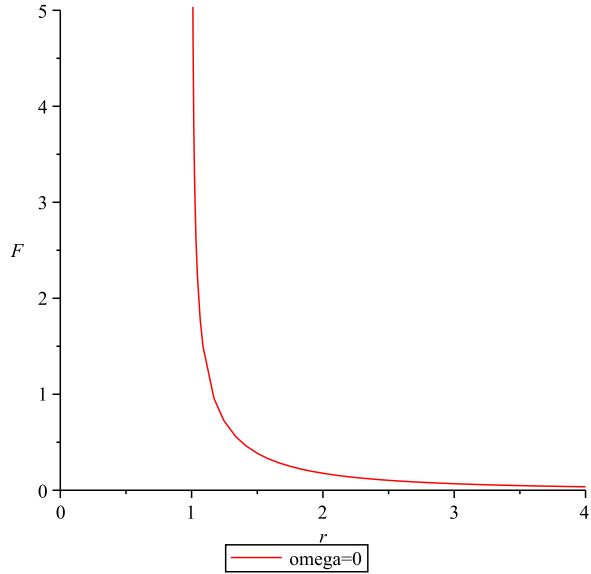} 
\label{figFstat} \caption{$F_{stat}\mbox{ for m}=\frac{1}{2}\mbox{, the force goes to infinity as }r \mbox{ goes to one}$.}
\end{figure}
We see that, as $r\rightarrow2m$, $\alpha\rightarrow0$ and both $F_{stat}$ and $F$ become infinite~(see fig.(1)). That is, as was already very well noted in \cite{Crawford1}, a \emph{static limit} must be defined. This is the region were the observer can remain at rest relative to any other observer in the asymptotically flat spacetime. It is also the region were circular accelerated orbits are possible. Therefore, no static or circularly orbiting observer can exist to measure the velocity outside the region defined by this static limit.

In this geometry we obtain for the static and the circular, (eq.~(\ref{vcirc-test})) observers the following velocity:
\begin{eqnarray}\label{schw-circ}
   v^2_{stat}&=& 1-\frac{\alpha}{E^2(1+´\omega^2r^2)}\,.
\end{eqnarray}
The static case is obtained, as was mentioned before, by setting $\omega=0$ in this expression. The fact that, $v\rightarrow1$ as $r\rightarrow2m$, is a consequence of the existence of the above mentioned static limit and not, any property of this geometry.

The radial and the spiral observers can be treated with a single expression eq.~(\ref{v-spiral}) which becomes:
\begin{equation}\label{v-spiral-schw}
    v^2=1-\frac{\alpha^2}{E^2_1E^2_2\left[1-\sqrt{1-\frac{\alpha}{E^2_1}}\sqrt{1-\left(\frac{L^2}{r^2}+1\right)\frac{\alpha}{E^2_2}}\right]^2}\,.
\end{equation}
We can for $\frac{\alpha}{E^2}\ll1$, use the following formula,
$$
    (1+\frac{\delta}{E^2})^{1/2}\approx1-\frac{1}{2}\frac{\delta}{E^2}\,,
$$
with $\delta=\alpha$ and, $\delta=\left(\frac{L^2}{r^2}+1\right)\alpha=\lambda\,\alpha$, in the first and second square roots of the denominator in eq.~(\ref{v-spiral-schw}) to obtain:
\begin{equation}
  v^2 = 1-\frac{4E^2_1E^2_2}{\left(E_2^2+\lambda E_1^2\right)^2}\,.
\end{equation}
This expression is independent of $\alpha$ and when $L=0$, we have $\lambda=1$, so that it becomes the expression for the square of the velocity measured by a radial geodesic observer \cite{Crawford1}. We notice that we always have $v<1$ except if $E_1$ or $E_2$ are zero or infinity.
\section{The de Sitter Schwarzschild case}
The de Sitter Schwarzschild spacetime, originally derived by Kottler has,
\begin{equation}\label{kot}
    V^2=\eta^{-2}=\beta=1-\frac{2m}{r}-\frac{\Lambda}{3}r^2\,,
\end{equation}
where $\Lambda$ is the \emph{cosmological constant}, which we will take to be positive.

There will be horizons for all $r_h$ such that $\beta(r_h)=0$. This is the equation for the roots of a third degree polynomial. We are interested in the range $r>0$, and so there will be two roots if (see Fig.~(2)):
\begin{equation}\label{Delta}
    0<\Lambda<\Lambda_{crit}\equiv\frac{1}{9m^2}\,.
\end{equation}
\begin{figure}\label{beta-fig}
\centering
\includegraphics[width=8cm]{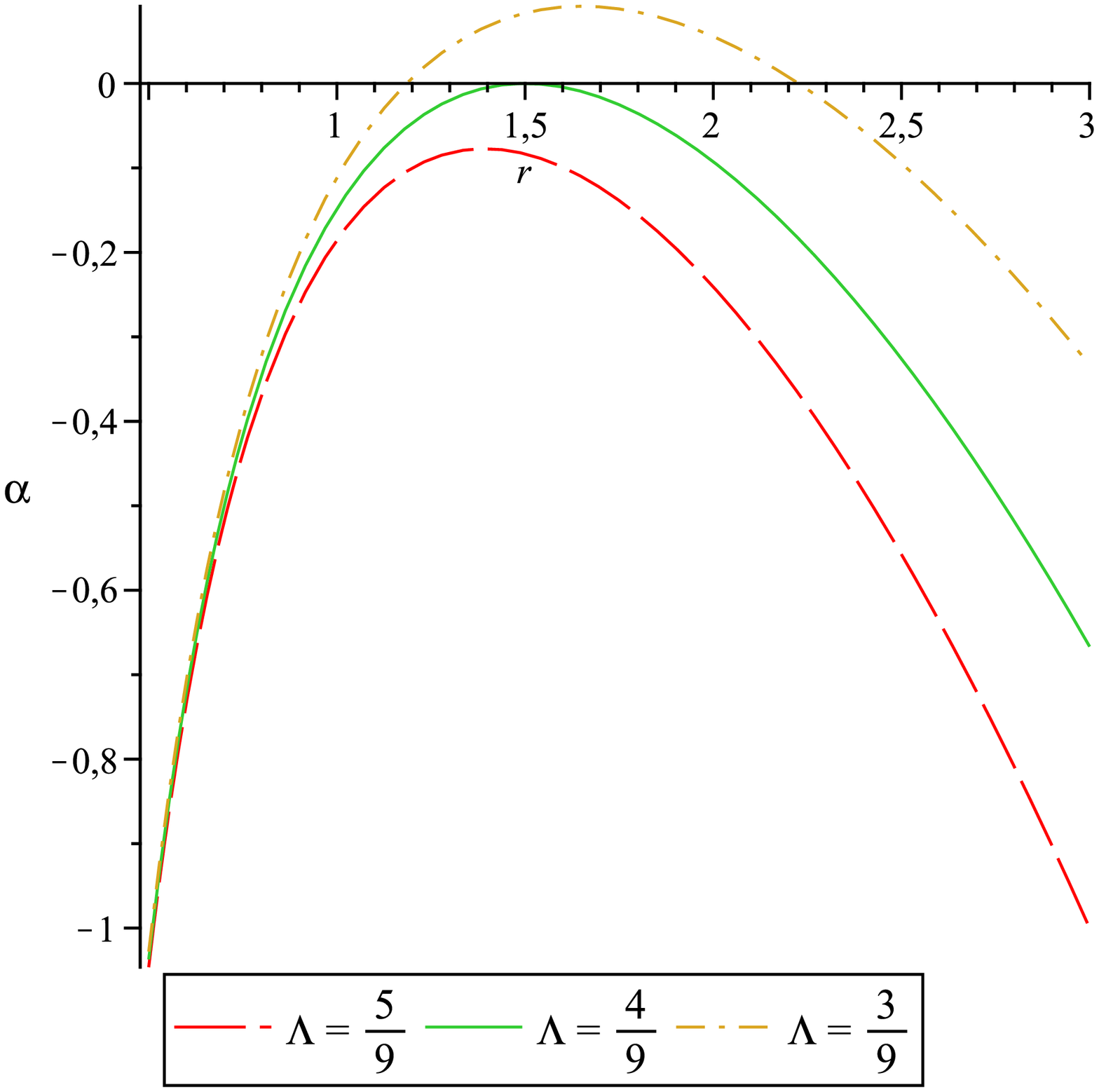} 
\label{figalpha} \caption{Plot of function $\beta$, for $m=\frac{1}{2}$} and three values of the cosmological constant. The first is $\frac{3}{9}<\Lambda_{crit}$ (dot dashed curve)
the second is $\Lambda_{crit}=\frac{4}{9}$ (solid curve) and and the third is $\frac{5}{9}>\Lambda_{crit}$ (dash curve).
\end{figure}
For $\Lambda>\frac{1}{9m^2}$, the spacetime is no longer static but homogeneous.\cite{cmc}.
For the range of $\Lambda$ given in~(\ref{Delta}) this geometry will therefore have two horizons.
This is in fact the combination of the Schwarzschild solution with it's event horizon at $r_S=2m$,
 with the de Sitter solution which has a cosmological event horizon at $r_{dS}=\sqrt{\frac{3}{\Lambda}}$ .
 They are, however different types of horizons. From the Schwarzschild horizon, nothing can exit.
 Whereas nothing can enter the de Sitter horizon. The first is a past horizon, and the second is a future horizon. The first encloses a singularity, but the second does not.
For the range given by eq.~(\ref{Delta}) the two positive roots are:
\begin{equation}\label{sols}
    r_k=2\,\frac{1}{\sqrt {{\Lambda}}}\sin \left[\frac{1}{3}\arccos\left( 3
\sqrt{\Lambda}m \right)+\frac{1}{6}\pi +\frac{2}{3}k\pi\right]\quad\mbox{for}\quad k=0,1.
\end{equation}
A series expansion of $r_{0,1}$ gives:
\begin{eqnarray}
  r_0 &=& {\sqrt{\frac {3}{\Lambda}}}-m-\frac{1}{2}\sqrt {3\Lambda}
{m}^{2}-\frac{4}{3}\Lambda{m}^{3}+O \left( {m}^{4} \right)\simeq r_{dS}-m\,,\\
  r_1&=& 2m+\frac{8}{3}\Lambda{m}^{3}+{\frac {32}{3}}{\Lambda}^{2}{m}^{5}+O
 \left( {m}^{6} \right) \simeq r_S+\frac{8}{3}\Lambda{m}^{3}\,.
\end{eqnarray}
And, putting these last two equations together we have:
\begin{equation}\label{trunc}
    r_S<r_1<r_0<r_{dS}\,.
\end{equation}
That is, the $r$-coordinate of the black hole horizon is greater the the Schwarzschild radius and,
the $r$-coordinate of the cosmological horizon is lesser the the de Sitter cosmological horizon.

We also have the curious result that the photon radius $r_\gamma$, is equal to that of the Schwarzschild case that is $$r_\gamma=3m\,.$$
\begin{figure}
\centering
\includegraphics[width=8cm]{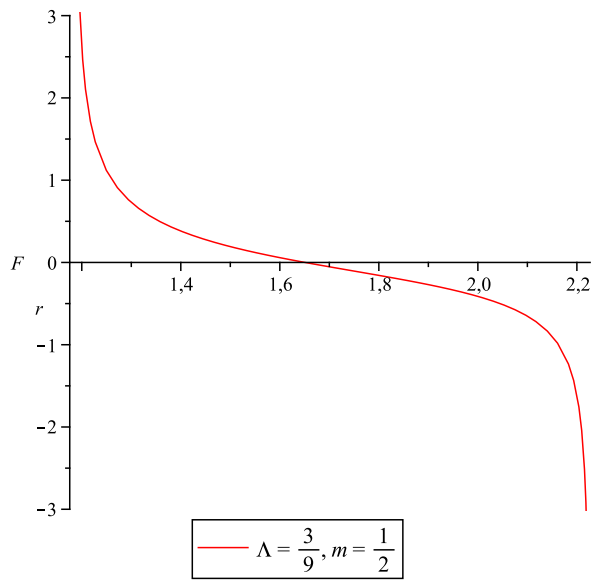} 
\caption{$F_{stat}$ for $\Lambda=\frac{3}{9}$ and $m=\frac{1}{2}$. The asymptotes are the roots of $\beta$ that is, $r_1=1.18479$ and $r_0=2.22668$.}
\end{figure}
The force needed to hold a unit mass stationary is:
\begin{equation}\label{Fskot}
    F_{stat}=\frac{\frac{m}{r^2}-\frac{\Lambda r}{3}}{\sqrt{1-\frac{2m}{r}-\frac{\Lambda r^2}{3}}}\,.
\end{equation}
This, will have two asymptotes at the roots of $\beta$, $r_{1,0}$, (see Fig.~(3)). The expression for $F_{stat}$, however approaches positive infinity in the first asymptote and negative infinity in the second. This is a sign that the horizons are different as we remarked above. For the black hole horizon a force must be exerted to keep an observer from falling in (a force that becomes infinite at the horizon), whereas for the cosmological horizon the force needs to be exerted to push the observer in, this force becomes infinite and the observer never passes through the horizon.

The Force for a circular observer with angular velocity $\omega$ is:
\begin{equation}\label{Fkot}
    F=F_{Stat} \left\{ 1-{\omega}^{2}r \left[ \frac{\beta}{\frac {m}{{r}^{2}}-\frac{1}{3}\Lambda\,r}-r \right]  \right\}
\end{equation}
As for the velocity measurements, they are similar to the Schwarzschild case but with the replacement $\alpha(r)\leftrightarrow\beta(r)$. There is, of course the obvious difference that, while $\alpha(r)$ only has one root, $\beta(r)$ has two, but the aforementioned nature of the cosmological horizon at $r_0$ makes velocity measurements impossible there.

In particular, the limit for the infalling observers is nominally the same but calculated at $r=r_1$ and not at $r=2m$, namely:
\begin{equation}\label{limit1}
  v^2(r=r_1) = 1-\frac{4E^2_1E^2_2}{\left(E_2^2+\lambda E_1^2\right)^2}\,.
\end{equation}
For the case $E_1=E_2$ we have
\begin{equation}\label{limit1}
  v^2(r=r_1,E_1=E_2) = 1-\frac{4}{\left(1+\lambda \right)^2}=1-\frac{4}{\left(2+\frac{L^2}{r^2} \right)^2}\,,
\end{equation}
since $\lambda=\frac{L^2}{r^2}+1$, and, a new approximation for $\frac{L^2}{r^2}\ll1$, yields
\begin{equation}\label{}
    v^2=\frac{L^2}{r^2}-\frac{3}{4}\frac{L^4}{r^4}+O\left( {\frac{L}{r}}\right)^6\simeq \frac{L^2}{r^2}\,.
\end{equation}
Thus, for equal energies and small rotational velocities, the measured velocity can be attributed to the radial motion of the observer and, the expression agrees with the Newtonian limit.

\section{Conclusions}

In this paper, we revisited a possible inconsistency related to the velocity measurement at a black hole event horizon. Previous methods had used equations (like (\ref{vstat}) and (\ref{schw-circ})) relating the velocity of a falling test particle to the $r$-coordinate of the trajectory it follows. For expressions such as these, the velocity goes the one as the test particle approaches the coordinate of the event horizon. One of the main misconceptions came from assuming this would happen with \emph{all observers}, and therefore that the velocity of a massive test particle become that of a photon, or in other words, that a timelike geodesic becomes a null geodesic.

Basing ourselves in a previous work \cite{Crawford1} and, aiming to further clarify this subject, we found not only that a static limit (a region where static observers may exist) must be defined in order to perform velocity measurements, but also that this limit seems to apply  to all accelerated observers as well.
If we (try to) use non-geodesic observers outside this limit, namely at the event horizon, the result is that the corresponding expression for the velocity has a limit equal to one. That is the case of the static and circular observers above. This nonphysical limit however is not a characteristic of the geometry but a reflection of the fact that we are using a nonphysical observer. This is depicted in Figures~(1)~and~(3) where the force need to hold such an observer is clearly seen to go to infinity. If on the other hand, we use geodesic observers, like the radial and spiral ones above, the limit of the velocity is strictly less then one, and no nonphysical measurements occur.

We developed general formulae for spherisymmetric spacetimes and applied to the Schwarzschild case and to the Schwarzschild de Sitter (Kottler) case. In this last case, the presence of a cosmological constant shifts the (black hole) event horizon in the sense of increasing $r$. However, qualitatively the behavior with respect to velocity measurements is the same. That is: accelerated (static and circular) observers can never be at the event horizon, and geodesic (radial and spiral) observers can exist and perform measurements there. They of course measure for the test particle a speed less then one.
\section{Acknowledgments}
The author gratefully acknowledges  Paulo Crawford for helpful discussions concerning this subject.

\end{document}